# Extracting useful information about reversible evolutionary processes from irreversible evolutionary accumulation models

Iain G. Johnston[1,*]

1. Department of Mathematics and Computational Biology Unit, University of Bergen, Bergen, Norway
* correspondence to iain.johnston@uib.no

Evolutionary accumulation models (EvAMs) are an emerging class of machine learning methods designed to infer the evolutionary pathways by which features are acquired. Applications include cancer evolution (accumulation of mutations), anti-microbial resistance (accumulation of drug resistances), genome evolution (organelle gene transfers), and more diverse themes in biology and beyond. Following these themes, many EvAMs assume that features are gained irreversibly – no loss of features can occur. Reversible approaches do exist but are often computationally (much) more demanding and statistically less stable. Our goal here is to explore whether useful information about evolutionary dynamics which are in reality reversible can be obtained from modelling approaches with an assumption of irreversibility. We identify, and use simulation studies to quantify, errors involved in neglecting reversible dynamics, and show the situations in which approximate results from tractable models can be informative and reliable. In particular, EvAM inferences about the relative orderings of acquisitions and the core dynamic structure of evolutionary pathways – which features are likely present when another is acquired – are robust to reversibility in many cases, while estimations of uncertainty and feature interactions are more error-prone.

## Introduction

Evolutionary accumulation models (EvAMs) describe the dynamics by which binary features are acquired by individuals over time (Diaz-Uriarte & Herrera-Nieto, 2022; Diaz-Uriarte & Johnston, 2025). Different approaches have been developed, largely independently, in the study of cancer progression (where features may correspond to particular mutations accumulating in a tumour (Beerenwinkel et al., 2015; Schill et al., 2020, 2024; Schwartz & Schäffer, 2017)) and evolutionary biology (where features may be genetic or phenotypic characters acquired over time in independent or related lineages (Boyko & Beaulieu, 2021; Johnston & Diaz-Uriarte, 2025; Lewis, 2001; Pagel, 1994, 1997)). Approaches described as EvAM have a particular focus on cases where features influence each other: for example, where the acquisition of one feature makes the acquisition of another more (or less) likely (Aga et al., 2024; Greenbury et al., 2020; Hjelm et al., 2006; Schill et al., 2020). Common targets of inference include a parameterised description of interactions between features, the likely orderings and timescales of feature acquisition, and/or transition graphs describing progress through different states of feature presence and absence (Fig. 1).

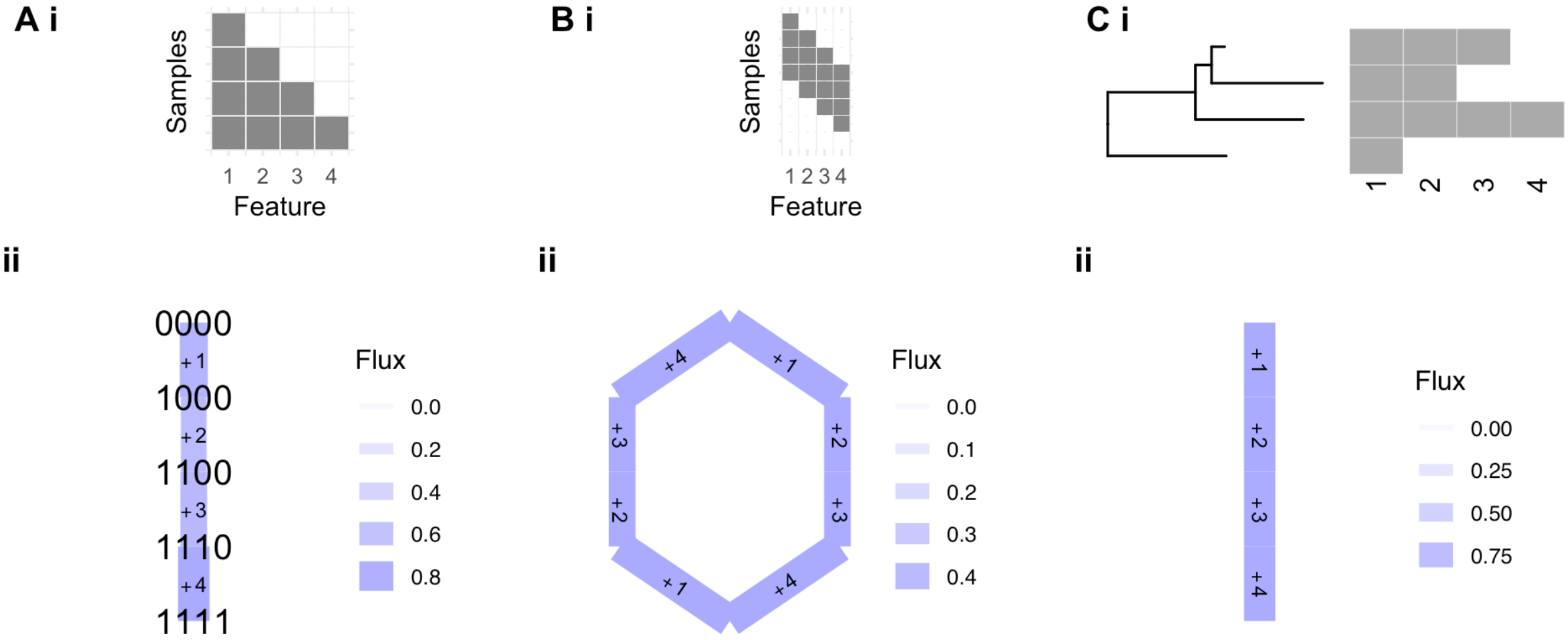

Figure 1. **Illustration of concepts in evolutionary accumulation modelling (EvAM). (A)** (i) Cross-sectional samples (rows) of a process involving the accumulation of 4 features (columns): grey indicates feature presence, white indicates absence. (ii) Transition network reflecting the accumulation process inferred from these data (using HyperHMM): starting from a state with no features acquired (0000), feature 1 is the first to be acquired, then feature 2, and so on. A single pathway is sufficient to account for all observations; different samples have proceeded to different extents down this pathway. The width of each edge gives the flux of probability through that edge. **(B)** A case involving two different pathways: accumulation ordered from 1 to 4 or, alternatively, from 4 to 1. Inference now assigns equal probability to these two competing pathways. In the transition graph (ii), node labels have been omitted; by convention the 0000 node is at the top of the plot and each vertical layer corresponds to the acquisition of one feature. These “competing” pathways require interactions between features: feature 1 represses the acquisition of feature 4 and vice versa. **(C)** A case involving phylogenetically linked data. Now, observations are not independent, and the EvAM process instead considers the transitions that occur down each independent lineage in the phylogeny (left).

EvAM approaches can be classified by the structure of data they analyse (Fig. 1). Many approaches assume that observations are cross-sectional – that is, completely independent instances of the accumulation process (Fig. 1A-B). For example, EvAM modelling in cancer often considers tumours in independent patients, which do not share a common ancestor from which mutations can be inherited. By contrast, EvAM in evolutionary biology (and, recently, in single-cell cancer biology (Aga et al., 2024; Luo et al., 2023)) often considers observations linked by a phylogeny (Fig. 1C). Here, the presence of a feature in an extant species may correspond to a recent acquisition, or inheritance from a common ancestor. In this evolutionary context, ancestral state reconstruction is an important part of the inference process (Aga et al., 2024; Johnston & Diaz-Uriarte, 2025; Renz, Dauda, et al., 2025).

Many evolutionary accumulation models assume irreversibility of changes: once a feature is acquired, it can never be lost. This assumption has two key consequences: *less constrained dynamics* and *uncertain ancestral state reconstruction*.

The first consequence, less constrained dynamics, influences all EvAMs. Irreversibility restricts the number of evolutionary pathways connecting two states to a finite set of paths on a hypercube, all of the same length. For example, the pathways from 000 to 011 are 000-001-011 and 000-010-011. Furthermore, once a target state is left, it will never be revisited. If transitions can be reversed, an infinite set of pathways can connect a source to a target state, and a path can visit a target state an infinite number of times (for example, 000-001-000-010-011-111-011-…) (Johnston & Diaz-Uriarte, 2025). If observations or assumptions constrain the timescale or other aspects of observation, these infinite possibilities will not all have equal probability. Pathways involving more steps will be less likely, and a likelihood function can at least in principle be constructed over a convergent set of such pathways. The simplest pathways – those corresponding to the irreversible accumulation picture – will likely constitute the highest-order terms in such a likelihood, but reversibility will certainly add complications.

The second consequence, uncertain ancestral state reconstruction, influences EvAMs where data is not cross-sectional. This includes phylogenetically or longitudinally related observations, which do not correspond to completely independent instances of the evolutionary process. Here, ancestral states must be considered to avoid pseudoreplication – considering coupled observations as if they are independent, and hence artificially increasing the effective sample size for some dynamics (Maddison & FitzJohn, 2015; Revell, 2010; Rohle, 2006). If dynamics are irreversible, reconstructing ancestral states is simpler (Dauda et al., 2025; Johnston & Williams, 2016; Renz, Brun, et al., 2025). If we further assume that evolutionary changes are rare (as is often assumed), the process of ancestral state reconstruction is deterministic: the most likely ancestral state is given by the AND operator applied over the set of descendant states (if and only if all descendants have a feature, the ancestor had it; otherwise it was acquired since the ancestor) (Aga et al., 2024; Greenbury et al., 2020; Johnston & Williams, 2016).

However, with reversible transitions, ancestral states cannot be deterministically derived from descendants. A distribution exists over the possible set of ancestral states, and this uncertainty challenges several EvAM approaches. Existing work has shown that for a variety of different cases, neglecting phylogenetic information influences the uncertainty but not necessarily the broad structure of evolutionary estimates (Dauda et al., 2025). Several EvAM approaches (Satas et al., 2020), including HyperMk (Johnston & Diaz-Uriarte, 2025), naturally account for uncertain ancestral states, but are limited in the number of coupled features they can analyse. Notably, HyperMk is an instance of the Mk model from comparative phylogenetics (Revell & Harmon, 2022), designed to support EvAM for reversible features. It is much more computationally demanding than approaches that assume irreversibility. However, it can also be parameterised to support only irreversible dynamics, making it a useful platform to compare irreversible and reversible models for accumulation dynamics.

Our goal here is to use simulated data, for which we control the true generative dynamics, to compare the performance EvAM approaches with irreversible and reversible dynamics. We are interested both in the ability of each approach to capture the true generating dynamics, and similarities and differences in the dynamics that the various models predict. From a practical perspective, we aim to characterise the situations where computationally tractable, irreversible models can provide useful information about dynamics which may have a reversible aspect.

## Methods

**Simulated data.** We begin by considering a class of generative evolutionary models that can capture the diversity of behaviours we want to investigate. We are interested in both irreversible and reversible "true" dynamics, in cases where a single dominant evolutionary pathway exists and cases where interactions between features create multiple competing evolutionary pathways. To this end, following previous work on EvAMs (Aga et al., 2024; Greenbury et al., 2020; Moen & Johnston, 2023; Williams et al., 2013), we consider two classes of "forward" dynamics: leftwards accumulation ...000-...001-...011 and rightwards accumulation 000...-100...-110... . For a single dominant pathway, we allow only rightwards accumulation (Fig. 1A); for competing pathways we allow both, with equal probability corresponding to the first step of each, but with each pathway repressing the other (Fig. 1B). We call these cases, involving explicit interactions between features enforcing orderings, "hard" pathways, due to their hard ordering constraints. We also consider independent features, where pathway structure is not enforced by explicit interactions, and all absent features have some possibility of acquisition from a given state. In all cases, each feature $i$ is acquired with a characteristic rate $\alpha_i$. For reversibility, we allow acquired each feature to be lost with a (different) characteristic rate $\beta_i$. Different choices of these rates can support "soft" pathways, where higher-rate features are usually (but not strictly) acquired first.

For each case study, following (Johnston & Diaz-Uriarte, 2025), we simulate a phylogeny using a birth-death model with birth rate 1 and death rate 0.1, using *ape* (Paradis & Schliep, 2019) in R (R Core Team, 2022). The number of tips is a parameter of the case study, as is $L$, the number of features that may be accumulated. On the phylogeny, we simulate character dynamics according to the chosen generative mechanism, starting from the state 000... at the tree root (as in Fig. 1C).

**Evolutionary accumulation models.** For $L \leq 5$, we use HyperMk to fit fully flexible reversible and irreversible models to the simulated data. We also use HyperHMM (Moen & Johnston, 2023) to analyse the data in two ways. First, we treat every observation (every tip of the phylogeny) as independent, and analyse the data as if it consisted of cross-sectional observations. Second, we reconstruct ancestral states deterministically as if feature acquisition was irreversible and rare, and use the corresponding transitions to fit a phylogenetically-informed HyperHMM model. We also use HyperTraPS (Aga et al., 2024; Greenbury et al., 2020) to estimate the interactions between features. We have previously shown that HyperHMM gives estimates of dynamics that are consistent with HyperTraPS in the case studies we consider (Aga et al., 2024; Moen & Johnston, 2023).

The different EvAMs consider different model structures (discrete vs continuous time, limited vs complete sets of possible interactions between features), but the dynamics they predict can readily be summarised in comparable ways. Our key summary will be the probability, under a fitted model, that feature $i$ is acquired when feature $j$ is absent (Dauda et al., 2025). For each fitted model, we construct a matrix $M$ where $M_{ij}$ is exactly this quantity, to facilitate comparisons.

**Software.** The code for this analysis is freely available at https://github.com/StochasticBiology/reversible-hyperinf . In addition to the libraries and programs above, it uses *dplyr* (Wickham et al., 2023) for data handing, and *ggplot2* (Wickham, 2016), *ggpubr* (Kassambara, 2020), *ggrepel* (Slowikowski, 2021), and *viridis* (Garnier et al., 2021) for visualisation.

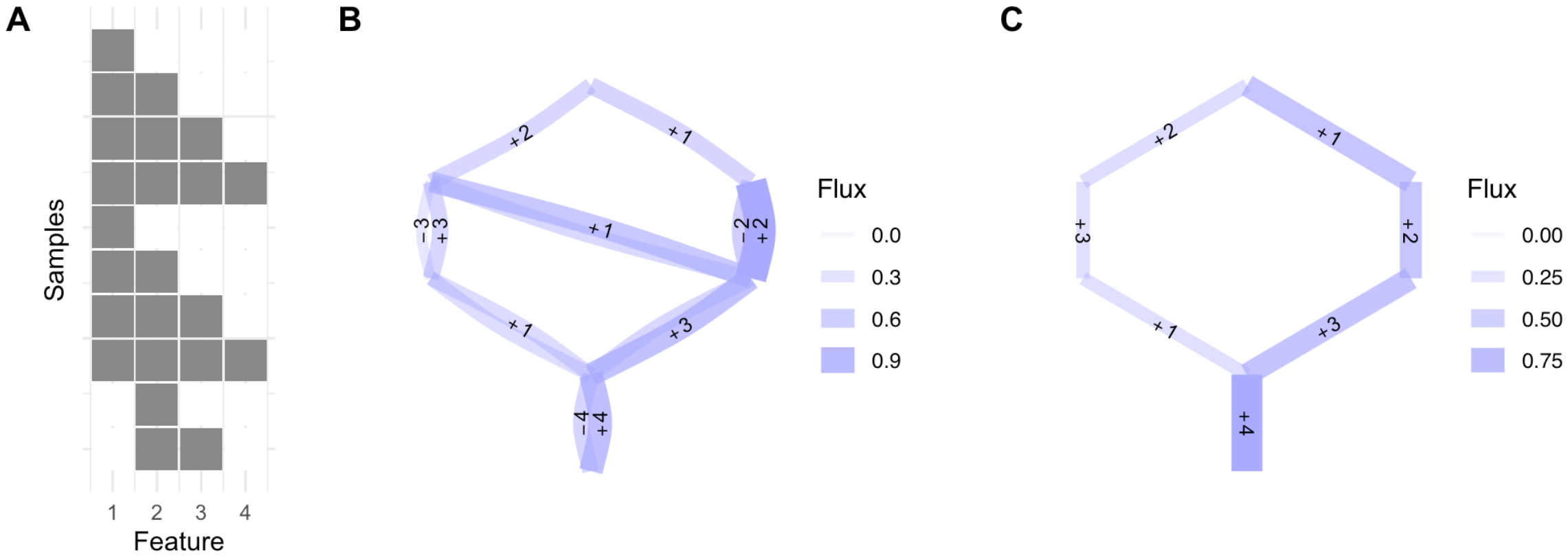

Figure 2. **Reversible and irreversible accumulation modelling in an extreme case. (A)** A dataset supporting a single pathway (acquisition ordering 1-2-3-4), with feature 1 capable of being lost once acquired. **(B)** Inference of likely accumulation dynamics from HyperMk, supporting reversibility (the double arcs in the figure correspond to forward and backward transitions). **(C)** Inference of likely accumulation dynamics from HyperHMM, assuming irreversibility. Except for the regain of feature 1 when lost, all high-probability acquisition transitions are captured (7/8 = 0.88).

## Results

**An extreme case of reversibility influencing inference**

The most extreme influence that reversibility can have on inferred dynamics occurs where one or more additional and independent irreversible pathways are required to account for a loss event. A simple example occurs when the data support one "hard" pathway (where interactions explicitly enforce orderings, see Methods): 000...**-**100... **-**110...-..., but an observation like 011... also exists. A reversible model can simply account for this through loss of the first feature, while an irreversible model needs an independent pathway.

To explore the information available from different inference approaches in such a case, we considered the cross-sectional dataset in Fig. 2A. Here, the first collection of observations support the hard pathway, with the final observations reflecting instances of states on that pathway where the first feature has been lost. It can readily be seen that the dynamics inferred by HyperMk (Fig. 2B), supporting reversibility, capture the loss of the first feature to provide a parsimonious explanation of the observations. However, the core dynamics of the reversible and irreversible (Fig. 2C) models are almost identical except for this transition. The orderings of acquisitions (typically 1-2-3-4, with some probability of 2 first) are preserved in both cases. This illustrates another important point: although reversible transitions may be possible in a model, it is not certain that pathways involving reversibility are always the highest-likelihood parameterisation for a given dataset.

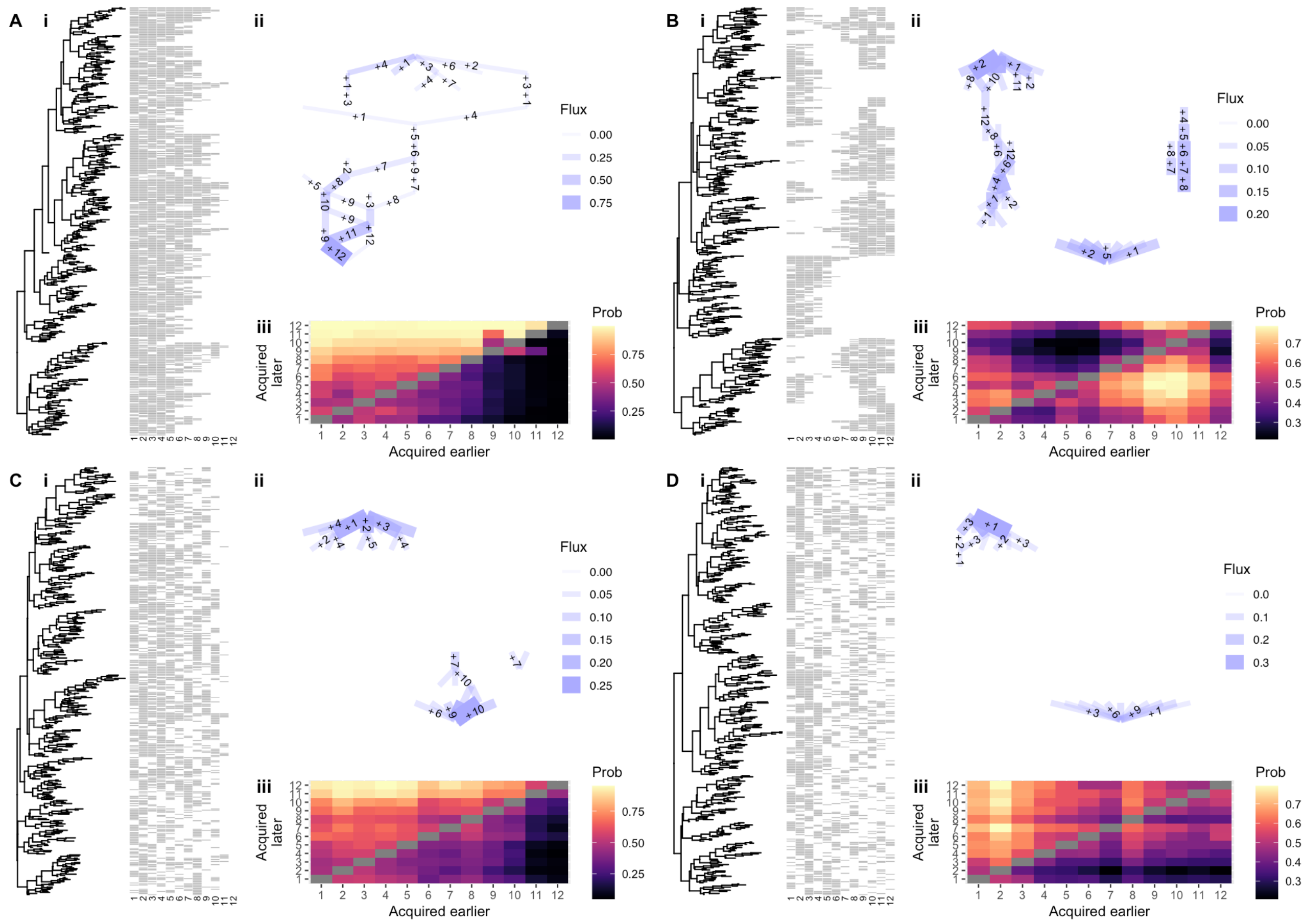

Figure 3. **Example inferences from diverse generative processes.** For each synthetic dataset (involving 40 phylogenetically-linked observations), a panel shows (i) the synthetic data; (ii) the highest-probability pathways inferred from HyperHMM; (iii) the summary matrix $M$ describing the probability that feature that feature $i$ is acquired when feature $j$ is absent. **(A)** One hard pathway with loss. While the pathway estimates in (ii) are noisier than the idealised case (for example, Fig. 1A), the core behaviour (lower-index features are acquired before later-index ones) is clear from the graph (ii) and ordering matrix (iii). **(B)** Two hard pathways with loss. Here, the idealised behaviour involves a uniform ordering matrix (each feature can be acquired early on one pathway and late on the other). Reversibility has made the pathway estimates less precise, and uneven sampling of the two pathways (i) has shifted inferred probabilities away from 0.5. **(C)** One soft pathway with loss. **(D)** One soft pathway induced by variable loss dynamics. Supp. Fig. 1 shows example inference outputs from (A-B) without loss dynamics.

### Diverse generative accumulation dynamics with and without reversibility.

To explore the impact of reversible transitions in larger systems, we simulated a collection of $L = 12$ dynamics with single and double pathways which may be "hard" (as above) or "soft" (where differences in rates give a typical ordering, but variability is possible because features are independent; see Methods), with and without reversible transitions. We used HyperHMM to infer evolutionary dynamics for different dataset sizes, both from a cross-sectional and a phylogenetically-coupled picture. To compare the inferred dynamics with known limiting cases, we consider a matrix $M$ describing the relative orderings of feature acquisitions. This is constructed for each inferred model as well as from theory describing the known generating mechanisms, and also from a collection of randomly reordered and assigned matrices. Some example results are shown in Fig. 3 and Supp. Fig. 1. We then compared these inferred outputs to various known cases, quantifying the number of shared edges in the inferred transition network (Fig. 4A) and using principal components analysis (PCA) to compare this collection of inferred dynamics in a low-dimensional space (Fig. 4B).

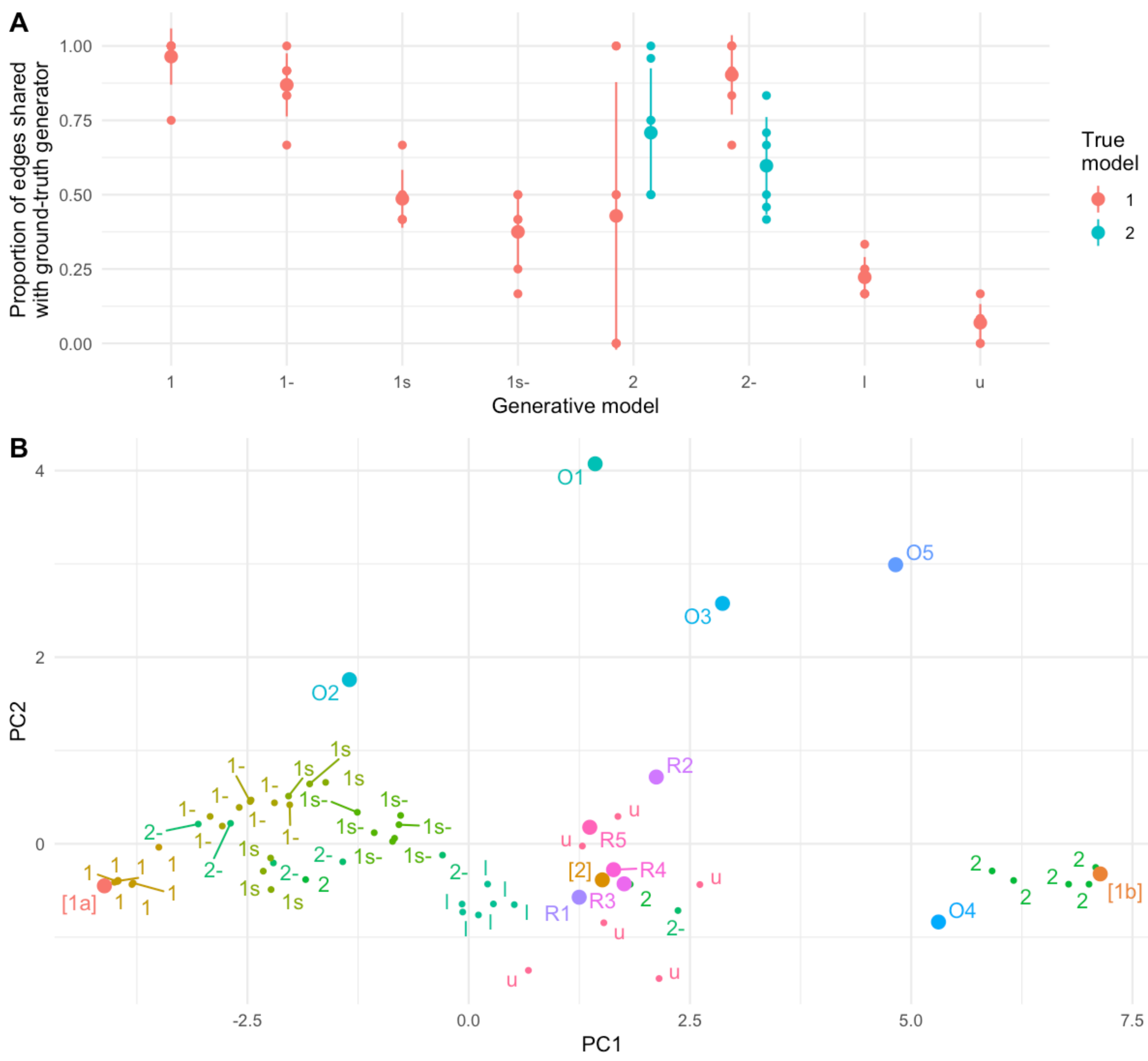

Figure 4. **Inference of evolutionary dynamics from diverse generative mechanisms. (A)** Proportion of transitions from the "ground truth" for hard pathways that are inferred with probability > 0.05 in the fitted models. **(B)** Principal components analysis (PCA) of ordering matrices $M$ describing the probability that feature that feature $i$ is acquired when feature $j$. Experiments: *1* (single hard path), *1-* (single hard path with loss), *1s* (single soft path), *1s-* (single soft path with loss), *l* (single soft path through variable loss rate), *2* (two hard paths), *2-* (two hard paths with loss), *u* (uniform acquisition). [1a] is the ground truth single path; [1b] is the ground truth for the opposite path; [2] is the ground truth for two paths. R1-5 are random matrices; O1-5 are randomly reordered variants of the single path. In this case, $L = 12$, data is phylogenetically coupled on a random tree, and there are 400 observations (as in Fig. 3). Supp. Fig. 2 shows outcomes from different experimental designs.

Several insights are visible from these observations. First, as a negative control, inferred dynamics for a process where all features are independent and identically acquired are random and uniform, with differences from sampling (label *u* in Fig. 4). As a positive control, the inferred solutions for the single, "hard", irreversible pathway (label *1* in Fig. 4) all cluster tightly with its ground-truth description and reproduce all ground-truth edges. We also consider the "soft" single pathway, where a likely but not certain ordering of acquisitions is imposed not by dependencies but by differences in independent acquisition rates (label *1s*).

Here, reflecting the increase variance in possible pathways, the solutions cluster (consistently) between the "hard" pathway and the uniform null model, with a given solution reproducing some but not all of the hard pathway's transitions.

The results for two "hard" irreversible pathways (label *2*) are more varied. In all cases, the inference detects the pathway structure. But for finite datasets, the weightings assigned to each pathway will differ in a given sample, and so we see differences in the probabilities in the summary ordering matrices as the dataset changes, with some but not all transitions captured. Larger datasets decrease this variability (Supp. Fig. 1; Supp. Fig. 2A).

All these observations are consistent with previous benchmarking of EvAM approaches on synthetic data (Aga et al., 2024; Greenbury et al., 2020; Moen & Johnston, 2023). We next consider the impact of reversible transitions. First, the presence of reversible transitions in the hard, single pathway picture (label *1-*) all cluster in a very similar position to the irreversible case, demonstrating that uniform loss dynamics do not compromise the inference of relative acquisition orderings. The proportion of ground-truth acquisition transitions captured remains high (75% or over) across different samples (Fig. 3A, Fig. 4). Reversible transitions in the "soft" pathway case (label *1s-*) lead to inferences similar to the soft pathway without reversibility, clustering similarly and with a similar comparison to the "hard" pathway statistics but, intuitively, with more spread of probabilities over alternative pathways (Fig. 3C, Fig. 4). Reversibility here has the effect of a shift in estimates towards the uniform case, reflecting a slight loss in precision for the pathway structure estimate.

In another picture, feature ordering is itself imposed by differences in the loss dynamics of different characters, with acquisition dynamics being the same (label *l*). Depending on parameterisation, the results of inference in this case do demonstrate the expected effective ordering, though with still less precision than the soft acquisition case (Fig. 3D, Fig. 4; see "Base rates and interactions" below).

Finally, reversible transitions in the case of two "hard" pathways (label *2-*) have a similar effect of decreasing the precision with which the two pathways are estimated, but inference of the core "bimodal" structure is retained (Fig. 3B, Fig. 4).

In all these cases, the effect of reversible transitions – either with uniform rates or with distinct rates for each feature – resembles that of increased noise in observations. As demonstrated above, this can lead to the EvAM process inferring additional pathways to account for the heterogeneous observations, in turn decreasing the precision with which ground-truth pathways can be estimated. However, the core pathway information is very rarely completely lost, and small sample size (so that the full dynamics are not adequately reflected in the sample) challenges inference substantially more than the presence of reversibility in these cases.

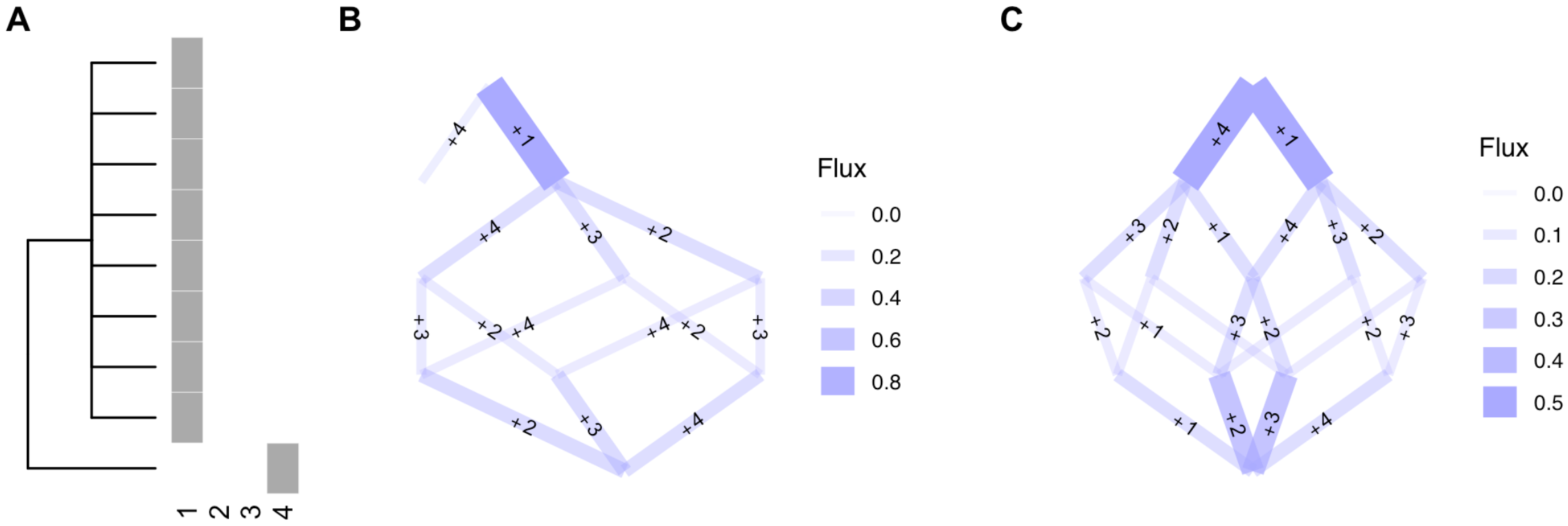

Figure 5. **Example of phylogenetic omission reweighting inference through pseudoreplication. (A)** A phylogenetic dataset with unbalanced representation of two lineages reflecting different dynamics. **(B)** EvAM output neglecting phylogenetic information. In assuming that each observation is independent, undue weight is given to the (repeated) observations of the feature-1-first pathway. **(C)** EvAM output including ancestral state reconstruction. Now the two pathways each have one effective representative and are equally weighted. The solution in (B) captures all the edges in (C), but because one pathway is assigned a low probability, only 14/20 = 0.7 of the edges appear with probability over a 0.05 threshold.

## Phylogenetic vs cross-sectional data

The inclusion or absence of phylogenetic information played little role in the point estimates from the various models (Supp. Fig. 2B). This limited effect mirrors previous observations (Dauda et al., 2025). The reason is that, in most conditions, the inclusion of phylogenetic information (including ancestral state reconstruction) serves mainly to limit pseudoreplication (treating related observations as if they are independent) (Maddison & FitzJohn, 2015; Revell, 2010; Rohle, 2006). Treating phylogenetically linked observations as independent effectively artificially elevates the sample size, which in turn leads to uncertainty estimates in EvAM being erroneously small. But it is unusual for the omission of phylogenetic information to have a substantial effect on the point estimates of the inference.

To demonstrate a case where such an effect does arise, consider the dataset in Fig. 5. Here, two pathways are supported: acquisition of feature 1 first, and acquisition of feature 4 first. One lineage follows the first pathway and produces 8 descendant observations. Another lineage follows the second pathway and involves only one descendant observation. If these observations are taken as independent, cross-sectional observations, we infer the first pathway to be roughly 8 times more likely. If ancestry is accounted for, we (under assumptions of rare and irreversible changes) reconstruct the common ancestor of the first lineage's observations as the only unique representative of that pathway, and the two pathways are given equal weighting in the inference.

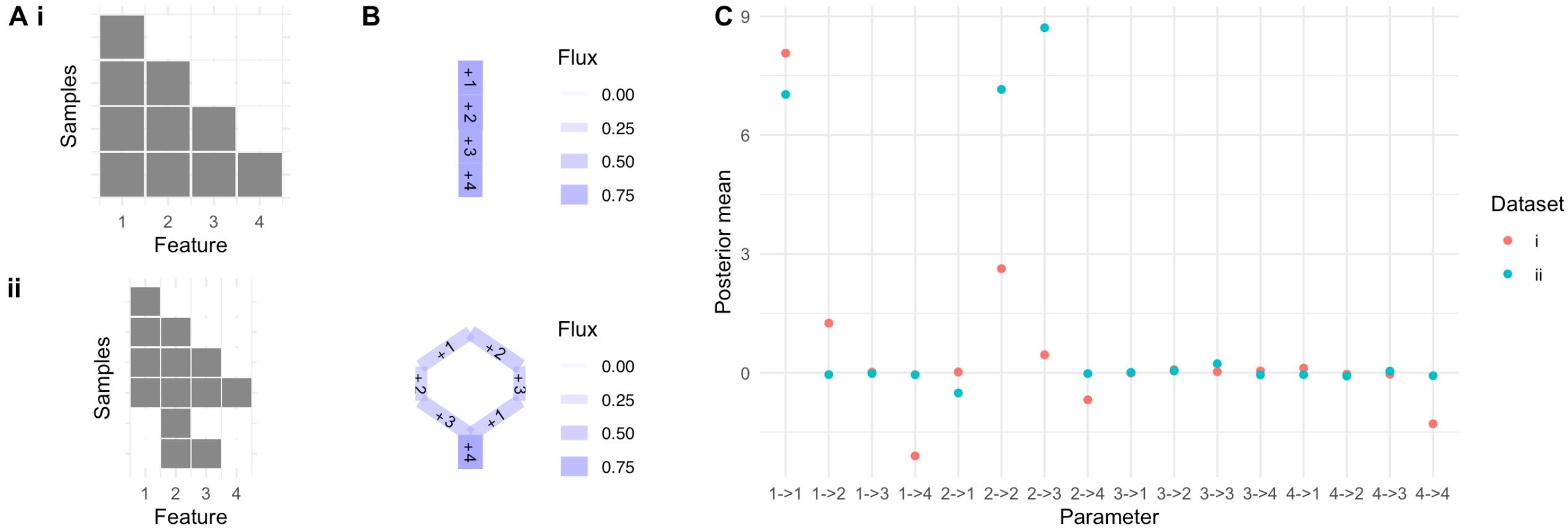

Figure 6. **Inferring interactions between features. (A)** Two datasets supporting a single path (i) and a single path with possible loss of feature 1 (ii), as in Figs. 1 and 2. **(B)** Inferred transition graphs using HyperTraPS (which assumes irreversibility) with penalised likelihood, matching the results from HyperHMM in Figs. 1 and 2. **(C)** Inferred interactions between features for the two datasets. Label $i \to j$ means the influence of acquiring feature $i$ on the base rate of acquiring feature $j$ ($i \to i$ gives the base rate for $i$).

This example illustrates that, in an extreme case, including or neglecting phylogenetic information can influence the relative weighting of different pathways but not naturally alter their structure. Outside of cases like this where there is a large imbalance in the dynamics subject to pseudoreplication, even this effect is limited. Given the generally limited effect of including any phylogenetic information, the errors resulting from an irreversibility assumption in the ancestral state reconstruction process will also often be limited, as observed in Fig. 4. However, as the uncertainty in inferences is directly linked to the effective sample size, uncertainty estimates will be more strongly influenced by the omission of phylogenetic information and potential pseudoreplication.

**Base rates and interactions.**

In the limit of independent features and long time scales, each feature's presence can be regarded as the long-term behaviour of a telegraph process. Its probability of presence will be $\alpha/(\alpha+\beta)$ and its likelihood contribution will be a binomial term with that probability. Hence, the target of inference for EvAM approaches will be $\alpha/(\alpha+\beta)$, suggesting that the inferred "base rate" of the feature's acquisition should be viewed as an estimate for this quantity. The observations in Fig. 4 for the soft pathway induced by variable loss (label $l$) reflect this, falling between the single-path and uniform regions in PCA space.

EvAM approaches like HyperTraPS (Aga et al., 2024; Greenbury et al., 2020; Johnston & Williams, 2016) and Mutual Hazard Networks (Luo et al., 2023; Schill et al., 2020) attempt to infer interactions between features – for example, whether the acquisition of feature 1 make the acquisition of feature 2 more or less likely. In reversible accumulation processes, feature presence could in principle also influence the loss propensity of other features – a collection of parameters which irreversible EvAM approaches have no way to infer. Also, the changes in pathway structure necessary for an irreversible model to account for reversible observations – illustrated in our first example – may invoke interactions between acquisition processes that are not present in the true generative process, as an artefact of attempting to account for reversible behaviour. To see this, consider again the simple cross-sectional dataset in Fig. 2A (also Fig. 6A). We now use HyperTraPS to infer the base rates and interactions between feature acquisitions. For the case without reversibility -- (i) in Fig. 6 – HyperTraPS assigns a high base rate to feature 1, a limited base rates to feature 2, a low base rate to feature 4, and positive interactions for feature 1 promoting feature 2 and feature 2 promoting feature 3, reflecting reasonable generating dynamics (Fig. 6C). However, when observations of the loss of feature 1 are included – (ii) in Fig. 6 – a very different set of base rates and interactions are inferred (Fig. 6C). Although (as in our first example) an evolutionary model is inferred capturing the true acquisition transitions, the inference of interactions between features is not reliable.

**Reversible and irreversible approaches applied to real-world data.**
To explore the congruence between reversible and irreversible EvAM approaches for real-world data, we consider a developing application domain for EvAM: the evolution of anti-microbial resistance (AMR) in bacteria (Aga et al., 2024; Dauda et al., 2025; Greenbury et al., 2020; Renz, Dauda, et al., 2025). Here, we use the CABBAGE database (Dickens et al., 2026) of AMR phenotypes across a range of bacteria. We focus on *Klebsiella pneumoniae* (Aga et al., 2025; Holt et al., 2015) resistance to the five drugs for which most observations exist in the database: gentamicin, ceftazidime, meropenem, ciprofloxacin, and trimethoprim-sulfamethoxazole. We apply HyperMk, HyperHMM, and HyperTraPS to infer the orderings of accumulation of drug resistance phenotypes, assuming a cross-sectional perspective (Fig. 7). In most cases, both the transition networks (Fig. 7B) and summary statistics of the ordering dynamics (Fig. 7C) are quantitatively highly similar, which we also observe across different random subsamples of these data (Supp. Fig. 3). The dominant difference (Supp. Fig. 3) is that reversible model fits sometimes support several initial steps which are then reverted by back transitions to recover the original state, leading to increased uncertainty in initial step posteriors but highly comparable dynamics thereafter.

## Discussion

Taken together, these case studies demonstrate a consistent picture: insights from irreversible models can capture some, but not all, features of reversible generative dynamics. The structure of core accumulation transitions in evolution pathways, and the relative ordering of feature acquisitions, usually faithfully reflect the true dynamics. Given these observations, we can generally phrase the outputs of these EvAM methods as reporting *which features are likely present when another is acquired*, which can be learned with some confidence. When counterexamples occur (like some instances of the two-path case in Fig. 4) it is less the presence of reversibility and more a limited sample size that challenges inferences. Except in extreme cases of unbalanced lineages, issues from neglecting reversibility in ancestral state reconstruction have little impact on these estimates. On the other hand, estimates of interactions between features, and uncertainty on all inferences, are challenged by neglecting reversibility.

It would be impossible to give a complete exploration of all possible generative mechanisms in a single manuscript. In focussing on “hard” and “soft” single and double pathways, we have attempted to capture most of the variability possible in such mechanisms: coupled and uncoupled features, promotion and repression between features, uniform and distinct loss rates, and single and distinct pathways, as well as a range of sample sizes, data structures, and system sizes. However, general results for the errors involved in neglecting reversibility will require a future theoretical complement to this simulation-based study.

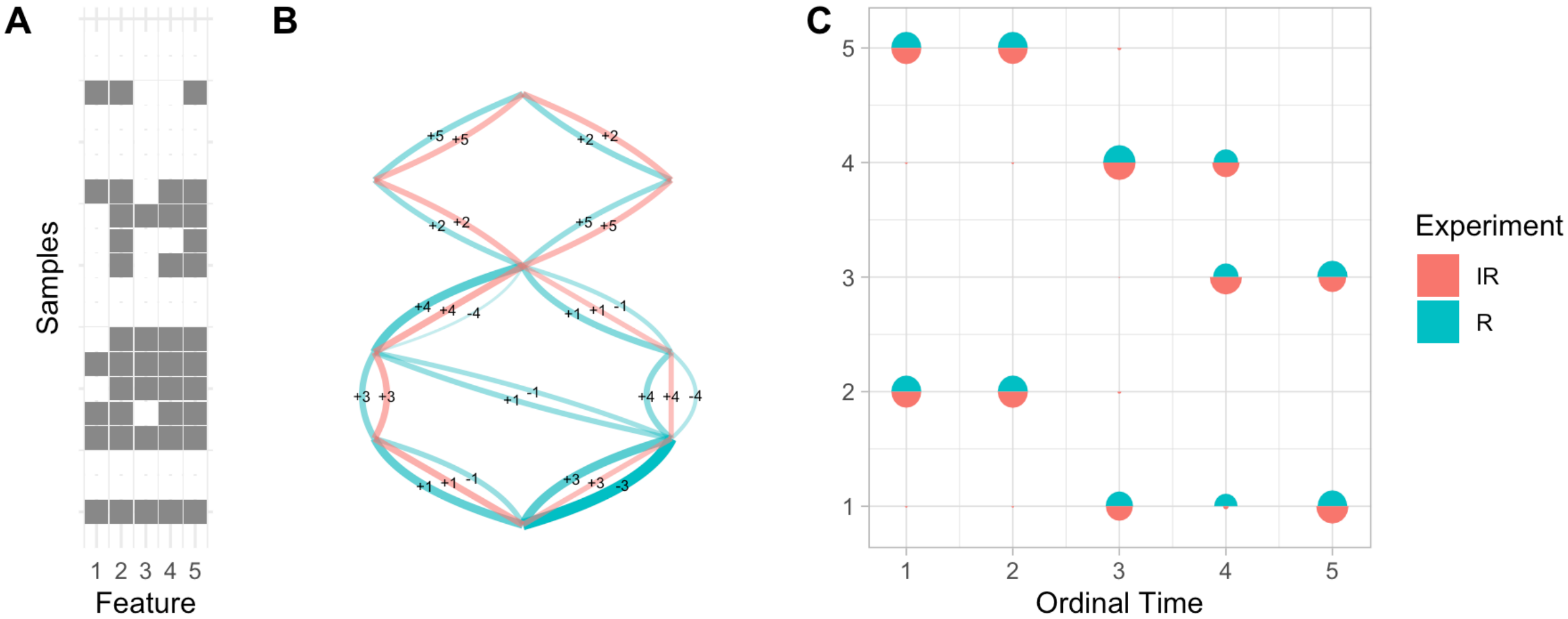

Figure 7. **Reversible and irreversible EvAM with AMR data. (A)** A subset of observations from the CABBAGE database (Dickens et al., 2026) of *Klebsiella pneumoniae* drug resistance phenotypes: dark pixels denote resistance to gentamicin (1), ceftazidime (2), meropenem (3), ciprofloxacin (4), and trimethoprim-sulfamethoxazole (5). **(B)** Transition networks inferred using reversible HyperMk ("R") and irreversible ("IR"). Reversible transition network can involve losses (-) as well as gains of features. **(C)** "Bubble plot" summary of dynamics. The size of a segment gives the probability that that feature (vertical axis) is acquired at that ordering step (horizontal axis) in the overall evolutionary process. Additional examples in Supp. Fig. 3.

An emerging use case of EvAM is in the study of multi-drug resistance (MDR) within the class of anti-microbial resistance (AMR) (Casali et al., 2014; Murray et al., 2022; Renz, Dauda, et al., 2025). Here, pathogens (like bacteria) evolve resistance to several of the drugs used to treat them. The features in EvAM are drug resistance phenotypes (or features of the genomes conferring these resistances), and the phylogenetically-related observations are bacterial genomes. In some bacteria, including *Mycobacterium tuberculosis*, MDR mutations are often acquired in the bacterial chromosome and relatively rarely lost (Aga et al., 2024; Casali et al., 2014; Greenbury et al., 2020). Some applications of EvAM in this context have neglected phylogenetic information and treated observations as independent (Leandry, Mujuni, Brun, et al., 2025; Leandry, Mujuni, Mureithi, et al., 2025). In other bacteria, including *Klebsiella pneumoniae*, MDR features are often acquired through horizontal gene transfer, where (for example) plasmids containing resistance genes may be lost readily after they are gained (Aga et al., 2025; Renz, Dauda, et al., 2025). The applicability of EvAM to such cases has remained an open question (Renz, Dauda, et al., 2025).

This study provides cause for cautious optimism in this context. We have shown that acquisition orderings can be estimated using irreversible EvAM even in the case where the true generating process involved reversibility (as is likely the case, for example, in the *Klebsiella* data in Fig. 7). While reversibility may lead to less clear results, inferring core aspects of the evolutionary pathways of reversible processes is possible and consistent with EvAM approaches based on irreversible models. The point estimates for relative feature orderings generally exhibit little bias, but estimates of uncertainty on these orderings and interactions between features are less reliable. Errors in ancestral state reconstruction, including those that arise from an irreversibility assumption, have limited influence on the outputs of inference – while the relative weightings of different evolutionary pathways may be changed, the core pathway structure remains intact.

# Supplementary Information

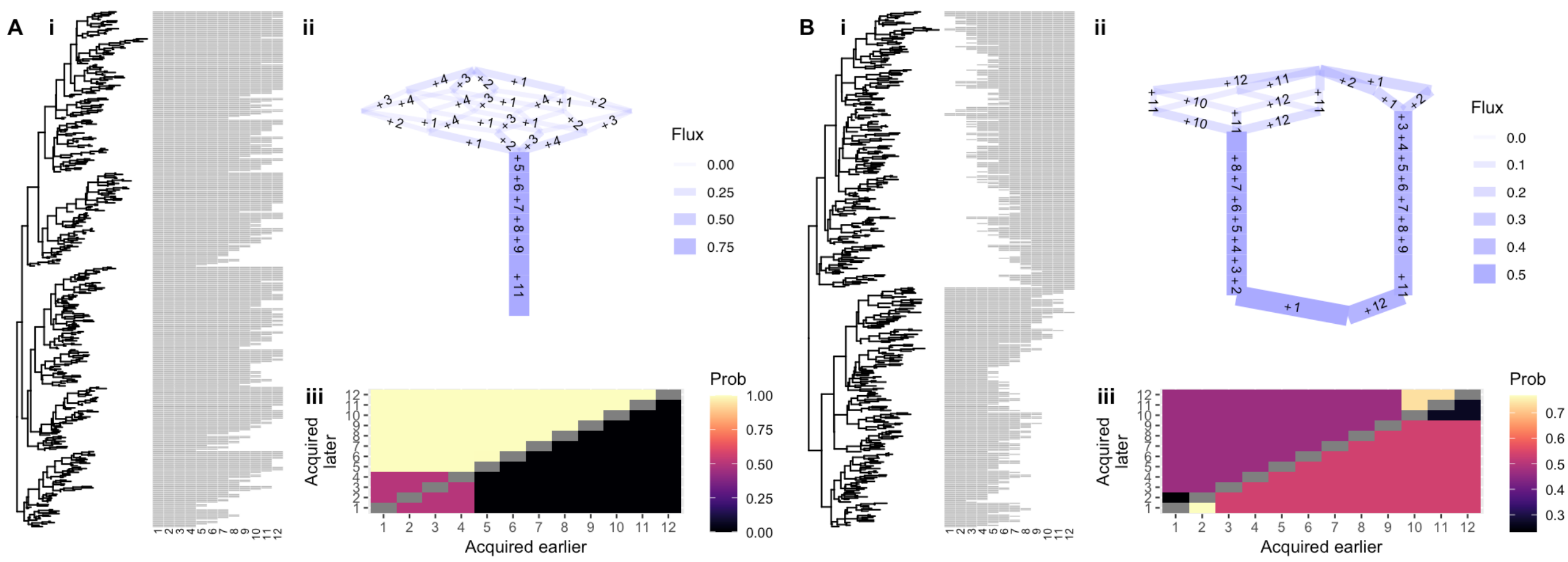

Supplementary Figure 1. Examples of inference of “hard” **(A)** single pathway dynamics and **(B)** two-pathway dynamics without reversibility (see Methods).

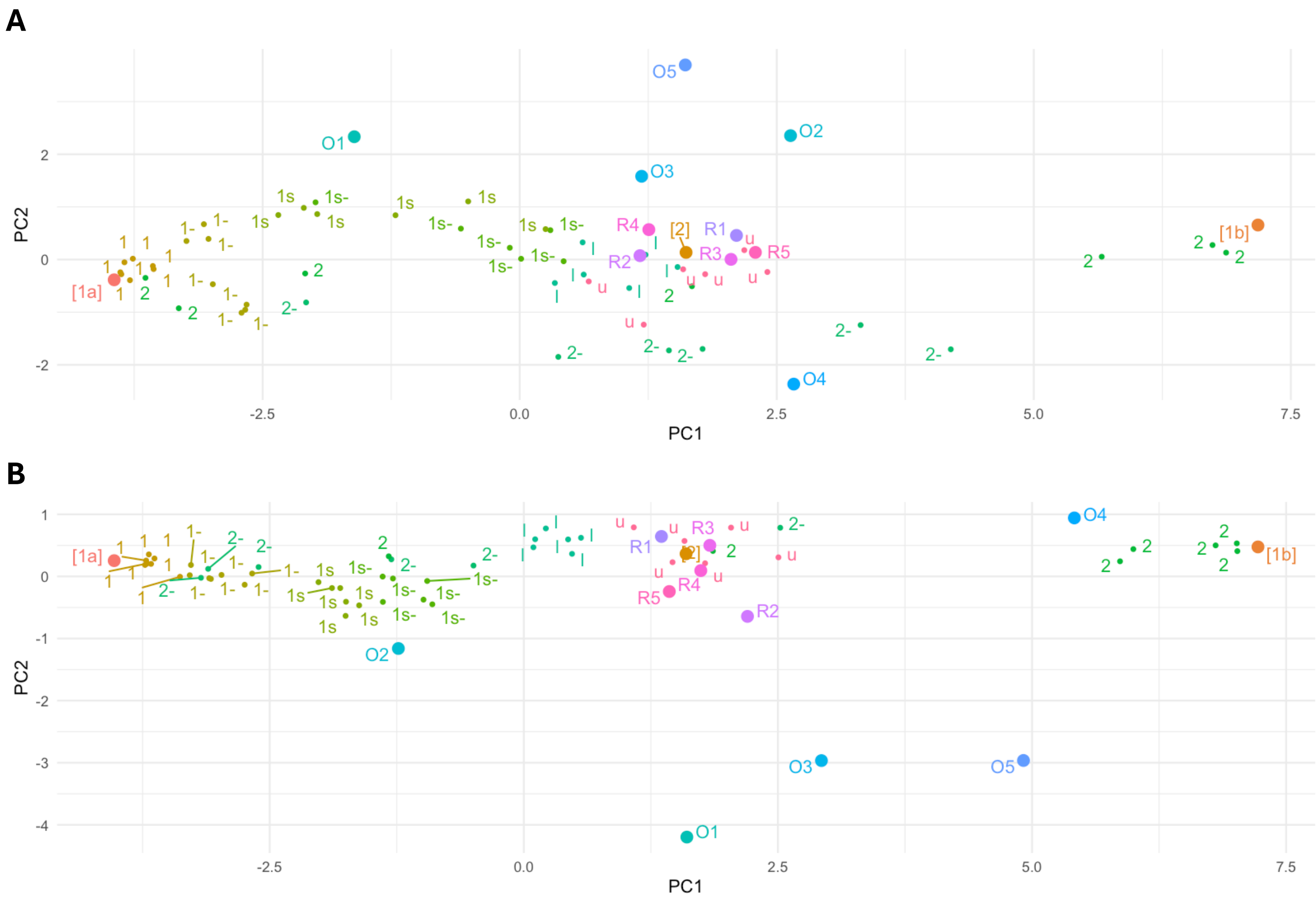

Supplementary Figure 2. **Inference of evolutionary dynamics from diverse generative mechanisms.** As in Fig. 4, plot shows ordering matrices $M$ describing the probability that feature that feature $i$ is acquired when feature $j$, compared via principal components analysis (PCA). Experiments: 1 (single hard path), 1- (single hard path with loss), 1s (single soft path), 1s- (single soft path with loss), l (single soft path through variable loss rate), 2 (two hard paths), 2- (two hard paths with loss), u (uniform acquisition). [1a] is the ground truth single path; [1b] is the ground truth for the opposite path; [2] is the ground truth for two paths. R1-5 are random matrices; O1-5 are randomly reordered variants of the single path. In Fig. 4, $L = 12$, data is phylogenetically coupled on a random tree, and there are 400 observations. Here, $L = 12$, **(A)** involves 40 cross-sectional observations; **(B)** involves the same data as Fig. 4 is inferred neglecting phylogenetic information. The randomised R1-5 and O1-5 are different in all cases.

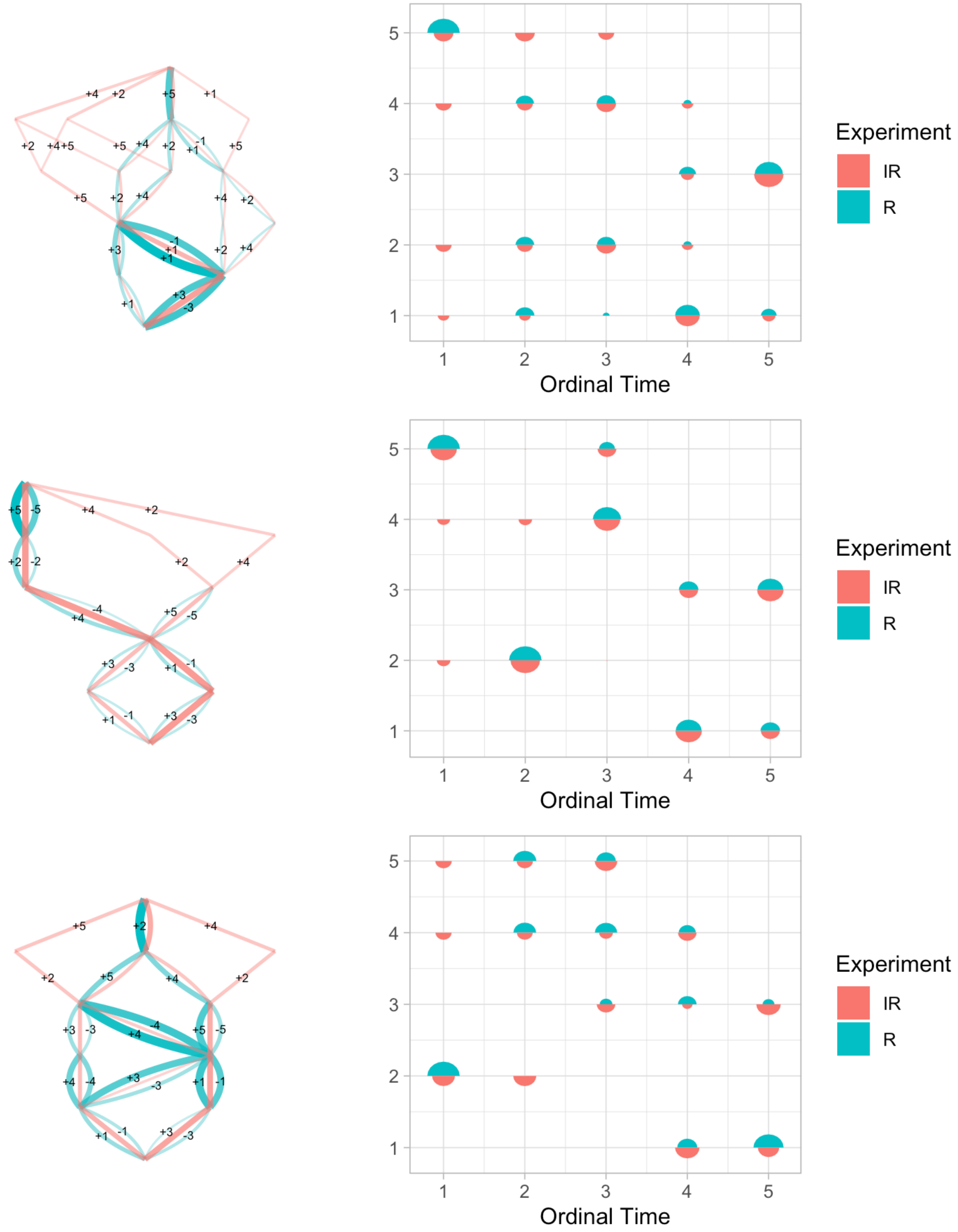

Supplementary Figure 3. **Consistency of reversible and irreversible approaches in different AMR samples.** Three additional examples of inferred dynamics of *Klebsiella* AMR evolution as in the main text: in each case the source data is a random subsample of the dataset. Here, the reversible fits allow some range of initial steps (increasing uncertainty) which are then reverted by subsequent transitions to recover the initial state; other transitions are all congruent across the approaches. In all cases, the Akaike Information Criterion (AIC) strongly favours the irreversible over the (more highly parameterised) reversible model fits (reversible AICs all > 500; irreversible AICs all < 400).